\documentclass[english,aps,preprint,amsfonts,amssymb,superscriptaddress]{revtex4}
\usepackage{amsmath}
\usepackage{babel}
\usepackage{bm}
\usepackage{enumerate}
\usepackage{graphics}
\usepackage{color}

\begin{document}

\newcommand\new[1]{\ensuremath{\blacktriangleright}#1\ensuremath{\blacktriangleleft}}
\newcommand\note[1]{$\blacktriangleright$[\emph{#1}]$\blacktriangleleft$}

\title{Light clocks in strong gravitational fields}
\author{Raffaele Punzi}
\email{raffaele.punzi@desy.de}
\affiliation{Zentrum f\"ur Mathematische Physik und II. Institut f\"ur Theoretische Physik, Universit\"at Hamburg, Luruper Chaussee 149, 22761 Hamburg, Germany}

\author{Frederic P. Schuller}
\email{fps@aei.mpg.de}
\affiliation{Max Planck Institut f\"ur Gravitationsphysik, Albert Einstein Institut, Am M\"uhlenberg 1, 14467 Potsdam, Germany}

\author{Mattias N.\,R. Wohlfarth}
\email{mattias.wohlfarth@desy.de}
\affiliation{Zentrum f\"ur Mathematische Physik und II. Institut f\"ur Theoretische Physik, Universit\"at Hamburg, Luruper Chaussee 149, 22761 Hamburg, Germany}

\begin{abstract}
We argue that the time measured by a light clock operating with photons rather than classical light requires a refinement of the standard clock postulate in general relativity. In the presence of a gravitational field, already the one-loop quantum corrections to classical Maxwell theory affect light propagation and the construction of observers' frames of reference. Carefully taking into account these kinematic effects, a concise geometric expression for the time shown by a light clock is obtained. This result has far-reaching implications for physics in strong gravitational fields.
 \end{abstract}

\preprint{AEI-2009-018}
\maketitle

\numberwithin{paragraph}{section}


The clock postulate in general relativity, which asserts that the time shown by a clock is proportional to the proper geometric length traversed along its worldline, is at the heart of the physical interpretation of the theory. It provides a raison d'\^etre for central theoretical concepts such as geodesic completeness used in the famous singularity theorems~\cite{Hawking:1969sw}, lies at the basis of virtually all conclusions we draw from currently collected astrophysical data~\cite{Spergel,Knop}, and plays an immediate role in the everyday application of the global positioning system~\cite{LachiezeRey:2006mj}. Inevitably, its relevance extends indirectly to all our fundamental matter theories built on the idea of a classical spacetime structure \cite{Wigner:1939cj}. The clock postulate is of central importance to all areas of relativistic physics.

Of course, the postulate can be physically relevant only if it agrees with the time shown by a physical clock made entirely of relativistic matter. Einstein clearly understood this and studied whether at least an idealized light clock, which operates by sending a classical light ray back and forth between two ideal mirrors held at a sufficiently small spatial distance, satisfies the clock postulate to any desired accuracy. Showing that this is the case, irrespective of the strength of the gravitational field, presents little more than a trivial exercise in Lorentzian geometry. The key point is that the entire workings of the clock can be derived from classical Maxwell theory.

However, the electromagnetic field is quantized, and this leaves an imprint on classical electromagnetic field configurations in the presence of a gravitational field. This has been calculated by Drummond and Hathrell for quantum electrodynamics~\cite{Drummond:1979pp}, leading to the one-loop effective action for~$F=dA$ on a vacuum spacetime~$g$, 
\begin{equation}
  W[A] \sim \int d^4x\, \sqrt{-g}\,\big(g^{a[c}g^{d]b} + \lambda C^{abcd}\big) F_{ab}F_{cd}\,.
\end{equation}   
The minute area scale $\lambda=\alpha/(90\pi m^2) \cong 3.85 \textrm{ fm}^2$ depends on the fine structure constant $\alpha$ and the electron mass $m$, and $C$ is the Weyl curvature tensor of~$g$.

How does this affect the time shown by a light clock? Clearly, the presence of the curvature term in the action $W[A]$ affects the entire causal structure of the electromagnetic field equations. Thus a  comprehensive discussion of a light clock in a strong gravitational field requires a revision of the construction of local observers' frames of reference from first principles. For while we are used to formally define central notions such as null, timelike and spacelike vectors directly in terms of the spacetime metric, it is well-known that their origin~\cite{EinsteinED}, and indeed physical relevance~\cite{Perlick},  lie in their role concerning the causal structure of fundamental matter equations, which in the classical domain are those of electrodynamics. Indeed, timelike covectors emerge as gradients to admissible initial data surfaces, timelike vectors as directions into which initial data can propagate, and the map between vectors and covectors is afforded by the duality map between the respective convex future cones~\cite{Hoermander,Benzoni}. It is a result of Maxwell theory, not a deeper principle, that these notions agree with their familiar characterizations in terms of a Lorentzian metric.  

In this letter, we provide a full answer as to how these kinematic constructions play out in the presence of strong gravitational field strengths up to the order of $\lambda^{-1}$, where classical electrodynamics is governed by the action $W[A]$. The geometric structure of this action becomes more apparent by writing it in the form of general linear electrodynamics~\cite{Hehlbook} for $G_{abcd} = 2 g_{a[c}g_{d]b} - \lambda C_{abcd}$,
\begin{equation}
  S[A]\sim\int d^4x |\textrm{Det } G|^{1/6} G^{abcd} F_{ab}F_{cd}\,.
\end{equation}
The geometric background $G$ is a covariant tensor with the symmetries $G_{abcd}=G_{[ab][cd]}=G_{[cd][ab]}$ whose determinant, defined by considering $G_{[ab][cd]}$ as a $6 \times 6$ matrix, is everywhere nonzero. Its contravariant inverse is defined by $G^{abmn} G_{mncd} = 4\delta^{[a}_c \delta^{b]}_d$. The following considerations will be based on the structure of $G$, not on its specific origin. In particular, employing the elegant language of cones, which play a noteworthy role throughout physics, we will exibit the causal properties of the theory. We will derive its timelike future cones and observers' frames of reference.  As a direct physical application, we will calculate that the time shown by a light clock in a background linearly perturbed away from a purely metric one is in accordance with a refined clock postulate. We will conclude with a discussion of the significance of this result, for classical gravity and beyond.

The analysis of the causal properties of the electromagnetic field equations obtained from the action $S[A]$  is essentially covered by the known theory of partial differential equations~\cite{Hoermander,Benzoni}, as we will demonstrate in more detail in~\cite{causal}. In particular, one finds that a pivotal role is played by the polynomial $P(k)=\mathcal{G}(k,k,k,k)$ defined in terms of the Fresnel tensor~\cite{Obukhov:2002xa,Hehl:2002hr,Rubilar:2007qm}
\begin{equation}\label{Fresnel}
   \mathcal{G}^{abcd} = \frac{|\textrm{Det }G^{-1C}|^{-1/3}}{-24}\epsilon_{mnpq}\epsilon_{rstu}G^{mnr(a}G^{b|ps|c}G^{d)qtu}\,
\end{equation}
where $\epsilon$ denotes the totally antisymmetric Levi--Civita symbol with $\epsilon_{0123}=1$, and $G^{-1C\,abcd}=G^{abcd}-G^{[abcd]}$. Up to an irrelevant factor, the polynomial $P(k)$ is the determinant of the principal symbol for the electrodynamic field equations on the background defined by the fourth-rank tensor $G$.  

Timelike covectors $k$ are defined in terms of this polynomial, since a hypersurface with gradient $k$ qualifies as a feasible initial data surface only if $P(k)\neq 0$ and if for any covector $q$ all roots of $P(q+\lambda k)$ are real. This is precisely the construction that leads to the familiar notion of timelike covectors in a Lorentzian spacetime; the important point here is that it effortlessly carries over to our more general case. Given a time-orientation of the manifold, by choice of an everywhere timelike covector field $t$, we further classify a timelike covector $k$ as future-pointing if all roots of $P(t-\lambda k)$ are positive. Clearly, the time orientation $t$ itself is future-pointing, and any other choice of a timelike covector field that is future-pointing with respect to $t$ induces the same classification. This is again precisely the same construction as in Lorentzian geometry. 

How do these notions translate from the cotangent to the tangent space? The set $\mathcal{T}^*$ of future-pointing timelike covectors at a point may look considerably more complicated than a familiar Lorentzian future cone, but it can be shown to be an open convex cone in the cotangent space~\cite{Garding}. But then there is a well-defined dual open convex cone ${\mathcal{T} = \{u \in T_pM\, |\,\forall k\in \mathcal{T}^*: k(u) \geq 0\}}$ in the tangent space, defining the future-pointing timelike vectors at a point. Indeed, the closure $\overline{\mathcal{T}}$ of this cone precisely contains the directions in which initial data can influence future field values. Note that the future cones $\mathcal{T}$ and $\mathcal{T}^*$ are perfectly determined by the tensor $G$ appearing in the action $S[A]$. Clearly, only a tensor field $G$ for which the timelike future cone $\mathcal{T}$ is non-empty at each point of the manifold may serve as a feasible background for electrodynamics. In technical terms, this makes the electromagnetic field equations weakly hyperbolic. In the special case of a metric geometry, precisely this criterion selects the Lorentzian metrics. There exists a non-linear bijective map from the future-pointing timelike covectors $k\in \mathcal{T}^*$ to the future-pointing timelike vectors $u \in \mathcal{T}$, which is given by
\begin{equation}\label{musical}
  u = \frac{\mathcal{G}(k,k,k,\cdot)}{\mathcal{G}(k,k,k,k)}\,.
\end{equation}
Invertibility of this map has been shown in the context of optimization theory~\cite{optimizationbook}. For the remainder of this paper, any pairs of covectors $k$ and vectors $u$ are assumed to be obtained from each other by the above map or its inverse, respectively.

The implications for relativistic kinematics are the following. For any observer with future-pointing timelike four-velocity $u\in \mathcal{T}$, there is a unique $3+1$ decomposition $T_pM = \langle u \rangle  \oplus V$ of each tangent space along the observer's worldline, into a time direction $\langle u \rangle$  and what we may call the spatial complement $V$, defined by $k(V)=0$. Additionally choosing a basis $(e_1,e_2,e_3)$ for $V$, the observer is equipped to perform local measurements of spacetime quantities.

What determines the geometry in the purely spatial directions seen by a particular observer? The background tensor $G$ has a spatial restriction to $V$ with components
\begin{equation}
  H_{\alpha\beta\gamma\delta} = G(e_\alpha,e_\beta,e_\gamma,e_\delta)\,
\end{equation}
with respect to the observer's local frame. Despite appearances, this really is a metric geometry; Cartan was probably the first to realize~\cite{Cartan} that in three dimensions any tensor with these symmetries is induced from a metric by virtue of $H_{\alpha\beta\gamma\delta}=h_{\alpha\gamma}h_{\beta\delta} - h_{\alpha\delta}h_{\beta\gamma}$, where the metric $h$ can be explicitly obtained~\cite{Punzi:2006nx} through 
\begin{equation}
  (\det h)\, h_{\alpha\beta} = \frac{1}{6} \epsilon^{\lambda\sigma\tau} \epsilon^{\rho\mu\nu} H_{\alpha\rho\sigma\tau} H_{\beta\lambda\mu\nu} \,.
\end{equation}
Thus a local observer sees a metric spatial geometry, even in the presence of a gravitational field. In particular, he may choose an orthonormal spatial frame such that $h(e_\alpha,e_\beta)=\delta_{\alpha\beta}$. Qualitatively, this of course coincides with what is always assumed. But quantitatively it differs, since the spatial metric $h$ is not merely the pull-back of a spacetime metric, but contains corrections from the spacetime curvature. And quite intriguingly, the spatial restriction of $G$ gives rise to a spatial metric only in spacetime dimension four.

Now we turn to the propagation of light rays. One finds from the geometric-optical limit of the electromagnetic field equations derived from $S[A]$ that light rays are described by the stationary curves $x$ of the action
\begin{equation}\label{lightaction}
  I[x] = \int d\lambda \, \mathcal{G}_{abcd}\dot x^a \dot x^b\dot x^c\dot x^d\,,
\end{equation} 
if also the null condition $\mathcal{G}_{abcd}\dot x^a \dot x^b  \dot x^c \dot x^d=0$ is satisfied everywhere along the curve~\cite{Punzi:2007di}. Note that it is not the Fresnel tensor $\mathcal{G}^{abcd}$ that features here, but a covariant dual Fresnel tensor $\mathcal{G}_{abcd}$. It is defined precisely like the Fresnel tensor (\ref{Fresnel}), but with all upper indices replaced by lower ones and vice versa, and $G^{-1}$ replaced by $G$. While this seems straightforward enough, it is actually a non-trivial exercise to prove that it is indeed the thus defined tensor that governs the action for light rays.   

\begin{figure}[h]
\begin{picture}(0,0)%
\includegraphics{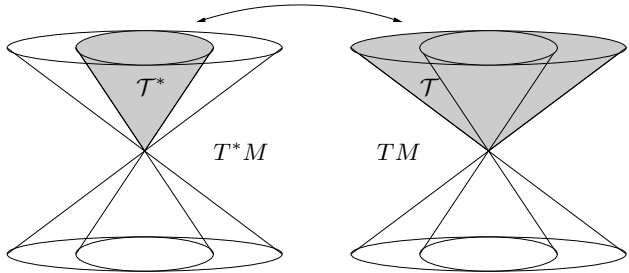}%
\end{picture}%
\setlength{\unitlength}{1421sp}%
\begingroup\makeatletter\ifx\SetFigFont\undefined%
\gdef\SetFigFont#1#2#3#4#5{%
  \reset@font\fontsize{#1}{#2pt}%
  \fontfamily{#3}\fontseries{#4}\fontshape{#5}%
  \selectfont}%
\fi\endgroup%
\begin{picture}(10824,4669)(2989,-4868)
\put(6601,-2911){\makebox(0,0)[lb]{\smash{{\SetFigFont{9}{10.8}{\familydefault}{\mddefault}{\updefault}{\color[rgb]{0,0,0}$T^*M$}%
}}}}
\put(9451,-2911){\makebox(0,0)[lb]{\smash{{\SetFigFont{9}{10.8}{\familydefault}{\mddefault}{\updefault}{\color[rgb]{0,0,0}$TM$}%
}}}}
\put(5251,-1786){\makebox(0,0)[lb]{\smash{{\SetFigFont{9}{10.8}{\familydefault}{\mddefault}{\updefault}{\color[rgb]{0,0,0}$\mathcal{T}^*$}%
}}}}
\put(10201,-1786){\makebox(0,0)[lb]{\smash{{\SetFigFont{9}{10.8}{\familydefault}{\mddefault}{\updefault}{\color[rgb]{0,0,0}$\mathcal{T}$}%
}}}}
\end{picture}%
\caption{\label{figure1}\textit{Perturbed timelike future cone $\mathcal{T}$ and its dual~$\mathcal{T}^*$.}}
\end{figure}

The emergence of the quartic null condition above, instead of the quadratic one familiar from Lorentzian geometry, deserves some comment. For a purely metric spacetime $G_{abcd}=g_{ac}g_{bd}-g_{ad}g_{bc}$, for instance as it underlies the action $W[A]$ in the absence of a gravitational field, one finds the dual Fresnel tensor $\mathcal{G}_{abcd}=g_{(ab}g_{cd)}$, and so one recovers the usual light cones. However, already for generic linear perturbations away from the purely metric case,  
\begin{equation}\label{linpert}
  G_{abcd} = 2 g_{a[c}g_{d]b} - L_{abcd}\,,
\end{equation}
one finds a less trivial dual Fresnel tensor
\begin{equation}\label{pertFresnel}
\mathcal{G}_{abcd} =  \Big(1+\frac{L}{6}\Big)g_{(ab}\left(g_{cd)}-L_{cd)}\right)+ \mathcal{O}\left(L^2\right),
\end{equation}
where $L_{ab}=L^{p}{}_{apb}$, $L=L^p{}_p$, and indices are raised and lowered using the metric $g$. Thus one obtains a null structure with two nearby lightcones, see figure~\ref{figure1}. In this case, different polarization components propagate according to one or the other of these two cones. So there is birefringence, and this needs to be taken into account when building a clock. For larger deviations from the metric case, even the bi-cone structure may break down, and a visualization of the situation is more difficult. But our general constructions of course cover this case equally well. 

With both the propagation of light and the local geometry of observers technically well under control, we are now in the position to investigate the physical validity of the following  
\begin{quote}{{\it Refined clock postulate:} The time measured by a clock with worldline $x(\tau)$ and future-pointing timelike tangent vector $u=dx/d\tau$ is proportional to $\tau$ provided the following normalization condition holds:
\begin{equation}
\mathcal{G}_{abcd}u^a u^b u^c u^d = 1 \,.
\end{equation}}
\end{quote}
Observe that the postulate reduces to the standard $(g_{ab}u^au^b)^2=1$ only in the absence of any gravitational forces, if the theory is applied to the action $W[A]$. That the conventional clock postulate nevertheless provides predictions in excellent agreement with experiments, such as the Pound-Rebka experiment~\cite{Pound:1960zz}, is due to the scale $\lambda^{-1}$, at which noticeable deviations appear, being so enormously much larger than the terrestrially encountered gravitational field strengths.

We will now investigate whether a light clock operating with photons ticks in accordance with the above postulate. For this purpose we need to set up an observer with an idealized light clock. Consider a Taylor expansion of $G$ around some point $p$, and an observer with future-pointing timelike velocity $u$ at this point. Assuming an idealized light clock that is smaller than the scale defined by the leading derivative terms of $G$, we may neglect all but the lowest order terms in a sufficiently small coordinate neighbourhood $x^a$ around $p$. It follows that the components $G_{abcd}$ can be taken as constant, that the observer is well approximated by the curve $x^a(\tau)=u^a \tau$, and that light rays in the chosen neighbourhood follow curves $x^a(\lambda)=z^a\lambda+w^a$ where the vector $z$ solves
\begin{equation}\label{nucon}
p(z)=\mathcal{G}(z,z,z,z)=0\,.
\end{equation}

The light clock now works as follows. The observer sends a light ray in the unit spatial direction $e\in V$, $h(e,e)=1$ in the observer's spatial complement $V$. The light ray hits an orthogonally placed mirror plane $\Omega\leq V$ at distance $D$ with $h(\Omega,e)=0$. This happens when
\begin{equation}
(e+\kappa_1u)\lambda = u\tau_1 + eD + \Omega\,,\quad p(e+\kappa_1u)=0\,.
\end{equation}
This is solved, if the mirror is centrally hit (which means selecting $0\in\Omega$), if $D=\lambda$ and $\tau_1=\kappa_1D$. The light ray is now reflected in spatial direction $-e$. The instance of the second reflection on a parallel mirror plane at zero distance constitutes a tick of the clock. This happens when
\begin{eqnarray}
&&(-e+\kappa_2u)(\lambda-D)+(e+\kappa_1u)D=u\tau_2+\Omega\,,\nonumber\\
&&p(-e+\kappa_2u)=0\,.
\end{eqnarray}
This is solved by $0\in\Omega$, $\lambda=2D$ and $\tau_2=(\kappa_1+\kappa_2)D$. The quantity $\tau_2$ is the period of a clock resolving both the direction of light $e$ and its polarization. We do not idealize the clock to this extent, but rather define a rotationally invariant clock through an average over $e$. We also average the polarization. Then the period of the resulting clock in the observer's parametrization can be expressed purely in terms of the mirror distance $D$ and the observer's direction $u$:
\begin{equation}
\tau_\textrm{clock}=\left<\kappa_1+\kappa_2\right>D\,. 
\end{equation}
It remains to calculate the averaged velocities $\kappa_1$ of the light ray and $\kappa_2$ of its reflection, depending on the particular background defined by $G$. Since this involves the explicit construction of all the required kinematic quantities, this is a formidable though now well-defined task in general.  

As an excellent check on our refined clock postulate, we may however comparatively easily calculate the time measured by a light clock for a geometry $G$ that deviates from its flat space value by a linear perturbation as in~(\ref{linpert}). First, we need to determine the three-space $V$ associated with a particular observer with future-pointing timelike tangent vector $u\in C$ at some spacetime point $p$. The isomorphism~(\ref{musical}) applied to $u$ yields the covector $k$,
\begin{equation}\label{cov}
  g^{ab}k_b = \frac{u^a}{g(u,u)} + \frac{L(u,u)u^a}{2g(u,u)^2}-\frac{L(u,dx^a)}{2g(u,u)}\,,
\end{equation}
where the last two terms provide the correction to the purely metric construction. Now choosing a basis $e_\alpha$ for the spatial complement $V$ defined by $k(V)=0$, one finds after some amount of linear algebra the three-dimensional Riemannian metric providing the spatial geometry seen by the observer:
\begin{equation}
  h_{\alpha\beta} = \Big(1+\frac{L}{4}-\frac{L(u,u)}{2g(u,u)}\Big) g_{\alpha\beta} - L_{\alpha\beta} + \frac{L(u,e_\alpha,u,e_\beta)}{g(u,u)}\,.
\end{equation}
This implies a useful result for vectors $e\in V$, $h(e,e)=1$:
\begin{equation}\label{useful}
\sqrt{g(e,e)} = 1-\frac{L}{8}+\frac{L(u,u)}{4g(u,u)}+\frac{L(e,e)}{2}-\frac{L(u,e,u,e)}{2g(u,u)}\,.
\end{equation} 
The light velocities $\kappa_1$ and $\kappa_2$ entering the clock period follow from $p(e+\kappa_1u)=0$ and $p(-e+\kappa_2u)=0$ for the polynomial $p$ defined in~(\ref{nucon}). We denote the two different polarizations by roman numbers; they each propagate along one of the cones defined by either one of the metrics $g$ or $g-L$ of the bimetric null structure~(\ref{pertFresnel}). We hence determine $\kappa_1^I$ from $g(e+\kappa_1^Iu,e+\kappa_1^Iu)=0$ and $\kappa_1^{II}$ from $(g-L)(e+\kappa_1^{II}u,e+\kappa_1^{II}u)=0$. We proceed similarly for the reflected light path to obtain $\kappa_2^I$ and $\kappa_2^{II}$ by replacing $e\mapsto -e$. Note that $k(e)=0$ and~(\ref{cov}) imply that $g(e,u)\sim \mathcal{O}(L)$. Then, using~(\ref{useful}),
\begin{eqnarray}
\kappa_{1/2}^I    &=& \mp \frac{L(e,u)}{2g(u,u)}+\frac{\sqrt{g(e,e)}}{\sqrt{-g(u,u)}}\,,\nonumber\\
\kappa_{1/2}^{II} &=& \kappa_{2/1}^I -\frac{L(u,u)}{2\sqrt{-g(u,u)}^3}-\frac{L(e,e)}{2\sqrt{-g(u,u)}}\,.
\end{eqnarray}
We now sum the light paths $\kappa_1+\kappa_2$ constituting one period, separately for each polarization $I$, $II$. The remaining dependence on the direction $e$ in which the light is sent is now quadratic. The average over $e$ can now be performed using the standard formula $\left<A(e,e)\right>=h^{\alpha\beta}A_{\alpha\beta}/3$. For all first order quantities $A$ this can be shown to equal
\begin{equation}
\left<A(e,e)\right>=\frac{A}{3}-\frac{A(u,u)}{3g(u,u)}\,, 
\end{equation}
where the trace is taken with $g$. Finally averaging also over the polarizations, we obtain the sought-for expression $\tau_\textrm{clock}$ for the worldline parameter describing one clock period
\begin{equation}
\tau_\textrm{clock} = \frac{2D}{\sqrt{-g(u,u)}}\Big(1-\frac{L}{24}+\frac{L(u,u)}{4g(u,u)}\Big) = \frac{2D}{\mathcal{G}(u,u,u,u)^{1/4}}\,,
\end{equation}
where the second equality follows from~(\ref{pertFresnel}). For this to be defined solely in terms of the clock properties (the linear length dimension $D$ is this case), and thus present an unambiguously measurable quantity, one needs that $\mathcal{G}(u,u,u,u)$ is constant. So indeed the light clock ticks in accordance with the refined clock postulate. 

\textit{Conclusions.} What we know about the structure of spacetime, we infer from observations of the matter inhabiting it. Abstract preconceptions about the spacetime geometry one entertains beyond that must be abandoned if observable matter proves them untenable (as the history of parity violation in particle physics may illustrate). In this letter we started from the observation that, as soon as the quantum character of the electromagnetic field is taken into account, the physical interpretation of the Lorentzian spacetime structure must be refined. The physical definition of kinematic notions such as timelike, spacelike and null vectors then deviates from their familiar characterization in terms of the spacetime metric. Most importantly, observers' frames of reference and the convex timelike future cones $\mathcal{T}$, defining the causal structure of the theory, are not those defined by the spacetime metric. 

These insights have direct implications for classical gravity and beyond. Microcausality in quantum field theory, for instance, can only be meaningfully defined with respect to the physical future cones $\overline{\mathcal{T}}$ identified in this letter, which nourishes the hope to reconcile it with the intriguing recent results on the high frequency behaviour of QED in curved spacetime by Hollowood and Shore~\cite{Hollowood:2007kt}. An illustration of the deep implications for classical gravity is provided by our explicit derivation of the readings of a photon clock: a physically meaningful definition of a complete spacetime must employ the physical proper time identified in the refined clock postulate. This renders some spacetimes incomplete that were previously classified as complete, and vice versa. This must be taken into account, for instance in the singularity theorems, if physical conclusions are to be derived.

The authors thank Gary Gibbons for valuable discussions and providing extensive notes on cones in physics. RP and MNRW gratefully acknowledge full financial support from the German Research Foundation DFG through the Emmy Noether grant WO 1447/1-1.


\end{document}